\documentclass[11pt]{article}
\evensidemargin=\oddsidemargin

\usepackage{graphicx}
\usepackage{epsfig}    
\usepackage{dcolumn}
\usepackage{bm}
\usepackage{amssymb}
\usepackage{epsfig}    
\usepackage{here}

\renewcommand{\thefootnote}{\fnsymbol{footnote}}

\def\beqn{\begin{eqnarray}}
\def\eeqn{\end{eqnarray}}
\relax
\newcommand{\ba}[1]{\begin{array}{#1}}
\def\ea{\end{array}}

\def\beq{\begin{equation}}
\def\eeq{\end{equation}}
\def\bea{\begin{array}}
\def\eea{\end{array}}

\def\to{\rightarrow}

\def\dis{\displaystyle}
\def\f{\frac}
\def\[{\left[}
\def\]{\right]}
\def\({\left(}
\def\){\right)}

\def\la{{\lambda}}
\def\ep{{\epsilon}}
\def\qs{{\sqrt{2}}}

\def\Ah{{\widehat{A}}}
\def\ms{{\widetilde{m}}}
\def\sm0{{\widetilde{m}_0}}
\def\sB{{\sin\beta}}

\def\tanB{{\tan\beta}}
\def\cotB{{\cot\beta}}
\def\sM{{\widetilde{\cal M}}}

\def\sL{{\widetilde{L}}}
\def\sE{{\widetilde{E}}}

\def\tas{\tilde{\tau}}
\def\mus{\tilde{\mu}}
\def\slp{\tilde{l}}

\def\cut{{\Lambda}}

\def\ov{\overline}

\def\U1em{{U(1)_{\rm em}}}
\def\to{\rightarrow}

\def\sq2{\sqrt{2}}

\def\End{\end{document}}
\newcommand{\lae}{\stackrel{<}{\sim}}
\newcommand{\gae}{\stackrel{>}{\sim}}

\begin{document}                                                              
\thispagestyle{empty}

\begin{flushright}
{\large hep-ph/0207030}
\end{flushright}
\vspace*{15mm}

\begin{center}
{\bf {\large
A More Flavored Higgs Boson in Supersymmetric Models
}}

\vspace*{15mm}

{\sc\large J. Lorenzo Diaz-Cruz}
\vspace*{4mm}  \\
~Instituto de Fisica, BUAP\\
Ap. Postal J-48, Puebla, Pue., 72570\\
 Mexico
\end{center}

\vspace*{55mm}
\begin{abstract}
\hspace*{-0.35cm}
A more flavored Higgs boson arises when the flavor
structure encoded in supersymmetric extensions of the
standard model is transmited to the Higgs sector.
The flavor-Higgs transmition mechanism
can have a radiative or mixing origin,
as it is illustrated with several examples,
and can produce interesting Higgs signatures that
can be probed at future high-energy colliders.
Within the minimal SUSY extension of the SM (MSSM), 
the flavor mediation mechanism is of {\it{radiative}} type,
as it is realized through gaugino-sfermion loops, and it transmit
the flavor structure of the soft-breaking sector to the
Higgs bosons; for this case we evaluate the
contributions from general trilinear A-terms to the 
Lepton Flavor-Violating (LFV) and Flavor-Conserving (LFC)
Higgs vertices.
On the other hand, as an example of flavor mediation through
{\it{mixing}}, we discuss an $E_6$-inspired multi-Higgs model,
suplemented with an abelian flavor symmetry, where  
LFV as well as LFC Higgs effects, are found to arise, 
though in this case at tree-level.
We find that Tevatron and LHC can provide information on the
flavor structure of these models through the detection of the
LFV Higgs mode $h\to \tau \mu $, while
NLC can perform high-precision tests
of the LFC mode $h \to \tau^+ \tau^-$.


\end{abstract}

\newpage
\setcounter{page}{1}
\setcounter{footnote}{0}
\renewcommand{\thefootnote}{\arabic{footnote}}

\noindent
{\bf {\large 1. Introduction.}}

One of the most important goals of future high-energy
colliders is to detect the Higgs boson, which remmains as
the only ingredient left to complete the Standard Model (SM),
and whose mass is constrained by SM radiative corrections 
to lay in the range 110-185 GeV at 95 \% c.l. \cite{hsmradc}.
A light higgs boson, with mass $m_h \lae 125$ GeV, is also 
predicted in weak-scale supersymmetry (SUSY) \cite{review}, 
which has become one of the leading candidates for physics
beyond the standard model. To be consistent with the
experimental data, supersymmetry has to be broken, i.e. 
the mass spectrum of the superpartners needs to be lifted.
 SUSY breaking is parametrized in the Minimal Supersymmetric SM 
(MSSM) by the soft-breaking largrangian, which preserves the
ultraviolet properties of exact SUSY.
In turn, the combined effects of the large top quark Yukawa coupling
and the soft-breaking masses, make possible to induce radiatively
the breaking of the electroweak symmetry. The Higgs sector of the
MSSM includes two higgs doublets, with the light Higgs boson being
perhaps the strongest prediction of the model.

 However, after a Higgs signal will  be seen, likely
at the ongoing or future hadron collider (Tevatron, LHC), it will become
crucial to meassure most of its properties, including its mass, spin and 
couplings, to elucidate its nature; this task is supposed to
be possible at the next-linear collider (NLC). In particular,  
the Higgs coupling to light fermions could be measured at NLC
with a precision of a few percent \cite{hpropnlc},
which will allow to constrain the
new physics laying beyond the SM.  For instance, higher-dimensional 
operators of the type  $\Phi^\dagger \Phi \bar{Q}_L \Phi b_R$
involving the third family,  will generate corrections
to the coupling $h\bar{b}b$, which in turn will
modify the dominant Higgs decay in the light mass range
\cite{ourtopeff};
NLC will have a chance to bound the strength of such operators. 
However, once we include the 3 families, those operators will
induce flavor-changing interactions in general, which can lead to a new
set of Higgs signals. Whether the diagonal or the off-diagonal terms
will play a more important role, will depend on the underlying
model that generates those operators, whereas its detection will
also depend on the capabilities of the different high-energy options
that are being consider by the high-energy physics community.

 The most widely studied scenarios for Higgs searches, assume that
the Flavor-Conserving (FC) Higgs-fermion couplings only depend on 
the diagonalized fermion mass matrices, while flavor-violating (FV) 
Higgs transitions are absent \cite{ourhixbb}.
However, when one goes beyond the minimal realization for these models, 
additional fields that do mediate FV transitions
often appear \cite{hifcnc}.
These new fields could also couple to the Higgs boson, either at
tree-level or radiatively, which in turn would induce Higgs-FV transitions.
In this paper we are interested in studying how the Higgs sector learns 
about the rich flavor structure encoded in SUSY models; focusing primarly
in the leptonic decays of the lightest Higgs boson.
As we shall argue below, the flavor-Higgs transmition can be classified
according to their origin, as {\it{ Radiative or Mixing}} mechanisms.  
 Flavor mediation through mixing could occur in general when extra 
bosons (fermions) mix with the SM bosons (fermions).

Within the MSSM, it can be shown that flavor-Higgs mediation
occurs through gaugino-sfermion loops, i.e. it is of radiative
type, and it communicates the non-trivial flavor
structure of the soft-breaking sector  to the  Higgs bosons. 
As an illustration of this case, we shall evaluate the 
SUSY contributions to the Higgs-lepton vertices,
including the slepton mixings coming from the
trilinear $A_l$-terms. The slepton mixing is 
constrained by the low-energy data, but it 
mainly suppress the FV's associated with the first two 
family sleptons, and still allows the flavor-mixings 
between the second- and third-family sleptons, the smuon ($\mus$) 
and stau ($\tas$), to be as large as $O(1)$~\cite{FCNC}. Thus,
in this scheeme one can neglect the mixing involving the 
selectrons, and the general $6\times 6$ slepton-mass-matrix 
reduces down to a $4\times 4$ matrix involving only the 
$\mus - \tas$ sector, similarly to the squark case
first discussed in ref. \cite{oursqmix}.  
Such pattern of large slepton mixing, can be motivated
by considering the large neutrino mixing detected with
atmospheric neutrinos, specially in the framework of
GUT models.

On the other hand, as an example of flavor-Higgs mediation through 
mixing,  we shall discuss an $E_6$-inspired  multi-Higgs model,
suplemented with an abelian flavor symmetry,
where large Higgs-FV  effects are also found to arise, though in this case 
at the tree-level. In this model there is a Higgs pair associated with 
each family. Then, to generate a realistic 
flavor structure for both leptons and sleptons \,\cite{FN,U1H},
we include a horizontal $U(1)_H$ symmetry, which
at the same time helps to keep under control the FCNC problem. 
Working in a basis where only one Higgs pair gets v.e.v.'s,
we prove that mixing effects between the light MSSM-like
Higgs boson and the heavier non-minimal states, induce
large tree-level corrections to the leptonic Higgs couplings.

The organization of this paper goes as follows: 
In Sect. 2 we discuss in general the possibilities for
having flavor-Higgs transmition, which includes the radiative and 
mixing cases. 
Among the models where flavor-higgs mediation occurs
through mixing, we shall discuss briefly the folowing cases:
i) the general two-Higgs doublet model (THDM),
ii) A model where the SM fermions mix with mirror fermions,
iii) Higgs-flavon mixing, and 
iv) R-parity breaking scenarios.
The radiative mechanism is discussed in detail in sect 3, within
the context of the MSSM with general trilinear soft-breaking terms.
This section includes the evaluation of the loop corrections both to the 
the lepton-flavor conserving (LFC) ($h \to l_i l_i$) and
the flavor violating (LFV) Higgs modes ($h\to l_i l_j $).
On the other hand, Sect. 4 includes the discussion of the
flavor-Higgs mediation within the context of
the $E_6$-inspired multi-Higgs model.
The phenomenological analysis for the capabilities of future
colliders to bound the Higgs-FC and -FV transitions that result
from this model is included in Sect. 4 too.
It is found that the induced Higgs-FV  couplings
can be significant enough to provide new discovery signals at the on-going 
Fermilab Tevatron Collider and the CERN Large Hadron Collider (LHC), 
which can detect the LFV mode $h\to \tau \mu $, 
and give information on the flavor structure of the model,
while NLC high-precision meassurements can bound the 
deviations from the SM for the LFC mode $h \to \tau^+ \tau^-$.
Finally our conclusions are presented in sect. 6.

\vspace*{6mm}
\noindent
{\bf {\large 2. Flavor-Higgs mediation mechanisms. 
}}  

Given the overwhelming experimental support for the SM
at present energies, and the indications of radiative
corrections favoring a light Higgs boson, it seems likely 
that this Higgs will be found sometime soon. Therefore, we 
can assume that the description of such light Higgs boson
will be given by an effective lagrangian, which
starts with the SM terms, but it includes
additional terms associated with new physics,
namely:
\beq
{\cal{L}}_{Higgs} = {\cal{L}}_{H_{SM}} + \Delta {\cal{L}}_{H}  ,
\label{eq:HiggsLag}
\eeq 
 ${\cal{L}}_{H_{SM}}$ includes the SM Higgs interactions
(Gauge, Yukawa and Higgs potential), whereas 
$\Delta {\cal{L}}_{H}$ denotes the correction to the Higgs
properties due to new physics.
The strength and structure of $\Delta {\cal{L}}_{H}$,
will depend on the nature of the new physics choosen by
nature at higher energies. For instance, it may include
the perturbative effects of heavy particles, after being integrated out,
or even be the remmaining manifestation of the underlying
mechanism of electroweak symmetry breaking  (EWSB). Very likely,
$\Delta {\cal{L}}_{H}$ will include corrections to the
interactions already present in the SM lagrangian, but it
could also include new interactions.

Within the SM, the Higgs boson-fermion couplings are only sensitive 
to the fermion mass eigenvalues. However, if one considers extensions 
of the SM, which either present a significant source of flavor-changing 
transition 
or are aimed precisely to explain the pattern of masses and mixing 
angles of the quarks and leptons, then it is quite possible that
such physics will include new flavor-mediating particles and interactions.
Furthermore, in the presence of additional fields that
have non-aligned couplings to the SM fermions, i.e.
which are not diagonalized by the same rotations that
diagonalize the fermion mass matrices, and that also couple
to the Higgs boson, then such fields could be responsible for
transmiting the rich structure of the flavor sector to the 
Higgs bosons interactions.

Depending on the nature of such new physics, we can 
identify two possibilities for flavor-Higgs mediation, namely:
\begin{enumerate}  
\item {\it{RADIATIVE MEDIATION.}} In this case the Higgs sector has 
initially (i.e. at tree-level) diagonal couplings to the fermions.
However, in the presence of new particles associated with extended flavor
physics, which couple both to the Higgs and to the SM fermions,
these flavor-mediating fields will induce corrections to the Yukawa
couplings and/or new FCNC proccess at loop level.
This case  will be discused  in great detail
in the forthcoming section, within the context of the MSSM with
general trilinear terms. 
\item {\it{MIXING MEDIATION.}} Modifications to the Higgs-flavor
structure can also arise when additional particles mix with
the SM ones. Such new particles could be either bosons or fermions.
 The possibility of having scalar flavor-mediation  arise when 
one considers new scalars with large FV couplings, 
these new interactions are then transmited to the Higgs 
sector, through scalar-Higgs mixing. Alternatively,
mixing of SM fermions with exotic ones could also
induce Higgs-FV couplings.
Both possibilities are illustrated next with several examples.
\end{enumerate}

 One model of Fermion-induced flavor-mediation was discussed 
in our previous paper \cite{mylrmmix}, where it was shown that
the presence of new mirror fermions can mix with the SM fermions,
and induce large corrections to the SM flavor structure; in particular
it allows for the presence of large Higgs-FV couplings. Besides
having to satisfy the low-energy constrains, it turns out that
these new interactions could be tested with the decay $h \to \tau\mu$. 
The importance of this LFV Higgs mode was presented in refs.
\cite{ourhlfv,prihmutau}, while subsequent work 
\cite{hanmarfat,mysnowtk,marcsher} further explored
the possibilities to detect it at future colliders.

The widely studied two-Higgs doublet model-III (THDM-III) could also
be considered as one case of scalar-induced flavor-Higgs mediation.
 In a basis where one Higgs doublet aquires a v.e.v.,
which resembles the SM Higgs, the fermionic couplings
of the second doublet will induce large FV transitions,
then through the mixing of both Higgs doublets, the light
SM-like Higgs boson will aquire such FV interactions
\cite{hifcnc,mythdiiia,mythdiiib}.

 Another example of scalar Flavor-Higgs mediation occurs when the 
flavon fields ($S$), 
which appear in the Froggart-Nielsen scheeme aimed to generate 
the hierarchy of fermion masses and mixing angles, mixes with
the light SM higgs. In some cases,
the flavor scale ($\Lambda$) can be close to the  electroweak-scale,
and the flavon $S$ could mix with the Higgs doublet ($\Phi$), for
instance  through a quartic term of the type
$S^\dagger S \Phi^\dagger \Phi$, which in turn will induce 
Higgs-FV couplings of the form: $(m_{l_i}+m_{l_j})/\Lambda$.
SUSY models with R-parity breaking provide another example of
scalar-induced flavor mediation; in this case
the sneutrino fields can aquire a v.e.v., which violates lepton number,
and this could be transmited to the Higgs sector by sneutrino-Higgs mixing.
Models where such mixing could appear were discussed some time ago in
Ref. \cite{RPValletal}.

 In sect. 5 of this paper, we shall discuss another model where  
scalar flavor-Higgs mediation is realized. It is an $E_6$-inspired
multi-Higgs model, where large LFV-Higgs  effects are also found to
arise, though in this case at the tree-level.
The model includes a Higgs pair associated with
each family, and is suplemented with an abelian flavor symmetry 
$U(1)_H$ that generates a realistic  flavor structure for both quarks
and leptons,
via proper powers of a single suppression factor\,\cite{FN,U1H}.
 This symmetry also helps to keep under control the usual FCNC
problem that appears in multi-Higgs models, whenever
each scalar field couples to both  u- and d-type fermions.
Working in a basis where only one Higgs pair gets v.e.v.'s,
we prove that mixing effects between the light MSSM-like
Higgs boson and the heavier non-minimal states, induce
large tree-level corrections to the leptonic Higgs couplings.

In summary: our previous discussion illustrates that the appearence 
of Higgs-FV couplings is quite a generic phenomena associated with
flavor physics beyond the SM, and in fact it can be used to
probe several aspects of the flavor problem.
 Although we shall explore the consequences of the
flavor-Higgs mediation mechanisms for the leptonic Higgs couplings, 
it can also be applied to the quark sector. 
In fact, implications for B-physics were 
discussed first in ref. \cite{babukolda},
mainly in the minimal SUSY-GUT context. Top quark physics was
discussed in our previous work \cite{oursqmix} for the MSSM,
where charged Higgs production through $cb$ fusion was
studied too. A more systematic evaluation of other rare top quark
decays \cite{mytcgama,mytcgamb}, will appear
elsewhere \cite{nextopfcnc}.

\vspace*{6mm}
\noindent
{\bf {\large 3. Flavor-Higgs transmition within the Minimal
Supersymmetric Model 
}}  

For good reasons, weak-scale supersymmetry
has become one of the leading candidates for physics beyond
the standard model (SM), notably
by sensibly explaining electroweak symmetry breaking (EWSB).
 Being a new fundamental space-time symmetry, SUSY
necessarily extends the SM flavor structure by
including superpartners for all fermions, and thus it
adds further puzzles to the flavor sector. 
 Within the Minimal Supersymmetric SM (MSSM)
SUSY is broken softly, in a manner that 
mantains its ultraviolet properties, while
respecting the phenomenological constraints. 
 However, the soft breaking sector of the MSSM is often
problematic with low-energy flavor changing neutral current (FCNC)
data without making specific assumptions about its free parameters. 
One of the most popular assumptions is the universality of squark
masses and proportionality of the trilinear $A$-terms
to the fermion Yukawa couplings. This is however not a generic
feature, and certain forms of non-diagonal $A$-terms
were studied recently \cite{nondA,Mura, oursqmix}.
Evolution from a high-energy scale, such as the GUT scale, is one 
possible source that generates non-minimal soft-breaking terms.
Moreover, in models that also attempt to
adress the flavor problem, the sfermion soft-terms may reflect
the underlying flavor symmetry of the fermion sector.
 The soft-terms flavor structure could then be transmited radiatively
to the Higgs sector, through gaugino-sfermion loops,
which is the focus of our paper. We shall evaluate here the 
corrections to the $h^0$ leptonic coupling, 
including both the lepton flavor violating (LFV) 
($h\to l_i l_j $) and the lepton-conserving (LFC)
($h \to l_i l_i$) decay modes.

\vspace*{3mm}
\noindent
{\bf 3.1. The MSSM with Non-Diagonal $A$-Terms.}  
\vspace*{3mm}

To evaluate the strength of the radiative Flavor-Higgs transmition
in the MSSM, we shall discuss first the form that the
slepton mass matrices and the Higgs-slepton and
gaugino-lepton-slepton interactions, take when expressed
in the sfermion mass eigenstate basis.

The MSSM soft-breaking slepton sector contains the following
quadratic mass-terms and trilinear $A$-terms:
\beq
\bea{l}
-\sL_i^\dag (M_{\sL}^{2})_{ij}\sE_j
-\sE_i^\dag (M_{\sE}^{2})_{ij}\sE_j  
+( A_{l}^{ij}\sL_i H_d\sE_j  + {\rm c.c.} )\,,
\eea
\label{eq:A-term}
\eeq
$\sL_i$ and $E_j$ denote the doublet and singlet
slepton fields, respectively,
with $i,j(=1,2,3)$ being the family indices.
For the charged slepton sector, this gives a generic $6\times6$ mass matrix,
\beq
\sM^2_u =\left\lgroup 
         \bea{ll}
          M_{LL}^2         &  M_{LR}^2\\[1.5mm]
          M_{LR}^{2\,\dag}   &  M_{RR}^2
         \eea
         \right\rgroup ,
\label{eq:MU6x6}
\eeq
where
\beq
\bea{ll}
M_{LL}^2 &= M_{\sL}^2+M_l^2+\f{1}{6}\cos2\beta \,(4m_w^2-m_z^2)\,, \\[2mm]
M_{RR}^2 &= M_{\sE}^2+M_l^2+\f{2}{3}\cos2\beta\sin^2\theta_w\, m_z^2\,, \\[1mm]
M_{LR}^2 &= \dis A_l v\,\sB/\sqrt{2}-M_l\,\mu\,\cotB \,,
\eea 
\label{eq:MU3x3}
\eeq
with $m_{w,z}$ denoting the masses of $(W^\pm,\,Z^0)$ and $M_l$ being 
the lepton mass matrix. For convenience, we will choose a basis where
 $M_l(=M_l^{\rm diag})$ is diagonal.

 In our {\it minimal} scheme, we consider all large LFV to
{\it solely} come from non-diagonal $A_l$ in the slepton-sector,
in a manner that respects the low-energy constrains, which
in fact allow the flavor-mixings between the smuon ($\mus$) and 
stau ($\tas$), to be as large as $O(1)$~\cite{FCNC}. 
Furthermore, such large mixing could be associated with the
large $\nu_\mu -\nu_\tau$ mixing observed in atmospheric
neutrinos \cite{superkam}.
 Thus, we can neglect the mixing between selectrons
and the other sleptons. In such minimal FCNC schemes
the general $6\times 6$ slepton-mass-matrix reduces down to
a $4\times 4$ matrix involving only the $\mus - \tas$ sector,
similarly to the quark sector discussed in ref. \cite{oursqmix}.
 Thus, we define at the weak scale,
\beq
A_l =
      \left\lgroup
      \bea{ccc}
      0 & 0 & 0\\
      0 & 0 & x\\
      0 & y & 1
      \eea
      \right\rgroup A_0 \,,
\label{eq:Al}
\eeq
where, $x$ and $y$ can be of $O(1)$, representing a naturally
large flavor-mixing in the $\mus - \tas$ sector. 
Moreover, identifying the non-diagonal $A_l$ as the only source of
the observable LFV phenomena implies
the slepton-mass-matrices $M_{\sL,\sE}^2$ 
in Eqs.\,(\ref{eq:MU6x6})-(\ref{eq:MU3x3})
to be nearly diagonal.  For simplicity, we define 
\beq
M_{LL}^2 \,\simeq\, M_{RR}^2\, \simeq\,\sm0^2\,{\bf I}_{3\times3}\,,
\label{eq:Degen}
\eeq
with $\sm0$ being a common scale for the scalar-masses\,\cite{seiberg}.

Within this minimal scheme, we observe that the first family sleptons
$\tilde{e}_{L,R}$ decouple from the rest in (\ref{eq:MU6x6}) 
so that, in the slepton basis $(\mus_L,\,\mus_R,\,\tas_L,\,\tas_R)$, 
the $6\times6$ mass-matrix is reduced to the following
$4\times4$ matrix,
\beq
\sM_{\slp}^2  =  
      \left\lgroup
      \bea{cccc}
      ~{\ms}_{0}^2~  &   0             &   0            &   A_x\\[1mm]
      0              &   ~\ms_{0}^2~   &   A_y~         &   0  \\[1mm]
      0              &   A_y~          &  ~\ms_{0}^2~   &   X_\tau\\[1mm]
      A_x~           &   0             &   X_\tau~         &  ~\ms_{0}^2
      \eea
      \right\rgroup 
\label{eq:Mct4x4}
\eeq
where
\beq
\bea{l}
A_x = x\Ah\,,~~~A_y = y\Ah\,,~~~\Ah = Av\,\sB/\sqrt{2}\,,~~~  
X_\tau = \Ah - \mu\,m_\tau\,\tanB \,.                     \\ 
\eea
\label{eq:MctADD}
\eeq
 The reduced slepton mass matrix (\ref{eq:Mct4x4}) 
has 6 zero-entries in total and 
is simple enough to allow an exact diagonalization.
 Therefore, when evaluating loop amplitudes one can
use the exact slepton mass-diagonalization, without
invoking the popular but crude mass-insertion approximation.

We have worked out the general diagonalization of 
(\ref{eq:Mct4x4}) for any $(x,\,y)$.
The mass-eigenvalues of 
the eigenstates  $(\mus_1,\,\mus_2,\,\tas_1,\,\tas_2)$ are,
\beq
\bea{ll}
M_{\mus1,2}^2 & =\sm0^2 \mp\f{1}{2}|\sqrt{\omega_+}-\sqrt{\omega_-}|\,, \\[2mm]
M_{\tas1,2}^2 & = \sm0^2 \mp\f{1}{2}|\sqrt{\omega_+}+\sqrt{\omega_-}|\,,
\eea
\label{eq:Mass}
\eeq
where $~\omega_\pm = X_\tau^2+(A_x\pm A_y)^2\,$. From (\ref{eq:Mass}), 
we can deduce the mass-spectrum of the smuon-stau sector as 
\beq
M_{\tas1} < M_{\mus1} < M_{\mus2} < M_{\tas2} \,.
\label{eq:Mspectrum}
\eeq
The $4\times4$ rotation matrix of the diagonalization
is given by, 
\beq
\bea{l}
\left\lgroup
\bea{l}
\mus_L\\
\mus_R\\
\tas_L\\
\tas_R
\eea
\right\rgroup
\!\!=\!\!
\left\lgroup
\bea{rrrr}
 c_1c_3  &  c_1s_3  & s_1s_4  &  s_1c_4  \\
-c_2s_3  &  c_2c_3  & s_2c_4  & -s_2s_4  \\
-s_1c_3  & -s_1s_3  & c_1s_4  &  c_1c_4  \\
 s_2s_3  & -s_2c_3  & c_2c_4  & -c_2s_4
\eea
\right\rgroup
\!\!
\left\lgroup
\bea{l}
\mus_1\\
\mus_2\\
\tas_1\\
\tas_2
\eea
\right\rgroup , 
\eea
\label{eq:rotation}
\eeq
with
\beq ~~~
s_{1,2}=\dis \f{1}{\qs} 
\[1-\f{X_t^2\mp A_x^2\pm A_y^2}{\sqrt{~\omega_+\omega_-}} \]^{1/2}\!,~~~
s_4
=\f{1}{\qs}\,,
\label{eq:rotation2}
\eeq
and $s_3=0$ (if $xy=0$), or, $s_3=1/\qs$
(if $xy\neq 0$), where $s_j^2+c_j^2=1$.

 Fig. 1 shows the resulting slepton spectra for typical
soft-breaking parameters in the range 0.3-0.9 TeV, $m_A=200$ GeV,
and $\tan\beta=5,20,40$. We can observe that both $\tas_1$ and
$\tas_2$ differ significantly from the common scalar mass $\sm0$.
Furthermore, the stau $\tas_1$ can be as light as about 
$100-300$\,GeV, which will have an important effect in the 
loop calculations. Furthermore, even for $x\simeq 0.5$ the smuon 
masses can differ from $\sm0$ by as much as 30-50 GeV; 
with these mass values the slepton phenomenolgy would have
to be reconsidered, since one is not allowed to sum over 
all the selectrons and smuons when evaluating slepton 
cross-sections, as it is usually assumed in the constrained 
MSSM.

\vspace*{5mm}
\noindent
{\bf 3.2. Higgs-sfermion and gaugino-sfermion interactions }
\vspace*{3mm}

To describe the radiative flavor-Higgs mixing, we need to
discuss the slepton-Higgs and gaugino vertices
in the mass-eigenstate basis. We shall present formulae
for the light Higgs boson ($h$), though the generalization
to the full Higgs spectrum is strightforward.
In terms of interaction states, the lagrangian that describes
the Higgs-slepton-slepton vertices can be written as:
\beq 
{\cal{L}}_{h\slp\slp} \, =\,  h^0[ \rho_L \slp^*_{Li} \slp_{Li}
 +\rho_R \slp^*_{Ri} \slp_{Ri} +(\rho_{LR} \slp^*_{Li} 
  \f{A_{ij}}{A_0} \slp_{Rj}+h.c.)]
\eeq
where:
\beq
\bea{ll}
{\rho}_L  &= g_Z m_Z \sin (\alpha+\beta) (\f{1}{2}+s^2_W) \\
\rho_{R}  &= g_Z m_Z \sin (\alpha+\beta) (-s^2_W)  \\
\rho_{LR} &=\f{ A_0 \sin\alpha}{\sq2} \,, 
\eea
\eeq
and $A_{ij}$ given by eq. (5).
Then, we can transform this lagrangian from the weak basis
$(\mus_L,\mus_R,\tas_L,\tas_R)$ to the mass-eigenstate
basis $(\mus_1, \mus_2, \tas_1, \tas_2)$.
The result can be written in terms of a $3\times 3$ rotated coupling
matrix: ${\cal{H}}_{\alpha \beta}={\cal H}_{\beta \alpha}=
                          O \tilde{\rho}_{\alpha\beta} O^T $,
where $\tilde{O}$ denotes the rotation matrix appearing in Eq.
(11), and
\beq
\tilde{\rho}  =  
      \left\lgroup
      \bea{cccc}
   \rho_L  &   0          &   0            & x \rho_{LR}  \\[1mm]
     0     &   \rho_{R}   &   y \rho_{LR}  &   0          \\[1mm]
     0     & y \rho_{LR}  &     \rho_{L}   &  \rho_{LR}   \\[1mm]
 x \rho_{LR}    &   0     &    \rho_{LR}   &  \rho_R
      \eea
      \right\rgroup 
\eeq

 The resulting Higgs-slepton couplings can be expressed then as:
  
\beq
{\cal{L}}= h^0 (\mus_1, \mus_2, \tas_1, \tas_2)
                {\cal{H}}_{\alpha \beta}
             (\mus_1, \mus_2, \tas_1, \tas_2)^T
\eeq

Similarly, the interaction between gauginos and lepton-slepton
pairs can be rotated to the mass-eigenstate basis too. The
result can be expressed as follows:
\beq
{\cal{L}}_{int}= \bar{\chi}^0_m
   [ \eta^{mL}_{\alpha k} P_L + \eta^{mR}_{\alpha k} P_R]
          \slp_\alpha  l_k +h.c.
\eeq
where $\chi^0_m$ denotes the neutralinos (m=1,..4), while $\slp_{\alpha}$
correspond to the mass-eigenstate sleptons. The factors
$\eta^{mN}_{\alpha k}$ are obtained after one substitute the
rotation matrices for both the neutralinos and sleptons
in the interaction lagrangian.

 To carry out the forthcoming analyisis of LFV Higgs transitions,
we choose to work with the simplified case $x=y$,
which gives: $c_1=c_2=c_{\slp}$, $s_1=s_2=s_{\slp}$ and
$c_3=s_3=c_4=s_4=\f{1}{\sq2}$.
 The corresponding expresion for the
matrix ${\cal{H}}_{\alpha\beta}$,
is given by,

\beq
\bea{ll}
 {\cal{H}}_{11}  &= \rho_L +\rho_R +2\rho_{LR} (s^2_{\slp}+ 
                    2 s_{\slp} c_{\slp} x) \\ 
 {\cal{H}}_{12}  &= (\rho_L -\rho_R) (s^2_{\slp}- c^2_{\slp}) \\ 
 {\cal{H}}_{13}  &= 2\rho_{LR} [s_{\slp} c_{\slp} - 
                    (s^2_{\slp}-c^2_{\slp})] \\ 
 {\cal{H}}_{14}  &= (\rho_L -\rho_R) 2s_{\slp} c_{\slp} \\ 
 {\cal{H}}_{22}  &= \rho_L +\rho_R +2\rho_{LR} (-s^2_{\slp}+
                    2 s_{\slp} c_{\slp} x) \\ 
 {\cal{H}}_{23}  &= (\rho_L -\rho_R) 2s_{\slp} c_{\slp} \\ 
 {\cal{H}}_{24}  &= 2\rho_{LR} [-s_{\slp} c_{\slp} - 
                    (s^2_{\slp}-c^2_{\slp})] \\ 
 {\cal{H}}_{33}  &= \rho_L +\rho_R +2\rho_{LR} (c^2_{\slp}-
                    2 s_{\slp} c_{\slp} x) \\ 
 {\cal{H}}_{34}  &= (\rho_L -\rho_R) (s^2_{\slp}- c^2_{\slp}) \\ 
 {\cal{H}}_{44}  &= \rho_L +\rho_R +2\rho_{LR} (-c^2_{\slp}-
                    2 s_{\slp} c_{\slp} x)  
\eea
\eeq

On the other hand, the expresions for $\eta^{mL,R}_{\alpha k}$,
simplify further for the case when the neutralino is taken
as the bino, which we will assume in the caculation of Higgs
LFV decays; the resulting coefficients ($\eta^{L,R}_{\alpha k}$)
are shown in table 1.

\newpage

\bigskip

\bigskip

\vspace*{5mm}
\noindent
{Table\,1. Slepton-lepton-neutralino couplings ($\eta^{mN}_{\alpha k}$)
for the case when $x=y$ and $\chi^0_1=\tilde{B}$. }
\vspace*{1.5mm}
\begin{center}
\begin{tabular}{|c|c|c|}
\hline\hline
$({\slp}_{\alpha},l_k)$ & $\eta^{L}_{\alpha k}$  &  $\eta^{R}_{\alpha k}$ \\  
\hline
$(\mus_1,\mu)$   &  $-c_{\slp} \f{g_1}{2}$
                 &  $-c_{\slp} g_1$           \\
\hline
$(\mus_1,\tau)$  &  $s_{\slp} \f{g_1}{2}$ 
                 &  $s_{\slp} g_1 $  \\
\hline
$(\mus_2,\mu)$   &  $-c_{\slp} \f{g_1}{2}$
                 &  $-c_{\slp} g_1$ \\
\hline
$(\mus_2,\tau)$  &  $s_{\slp} \f{g_1}{2}$ 
                 &  $-s_{\slp} g_1$   \\
\hline
$(\tas_1,\mu)$   &  $-s_{\slp} \f{g_1}{2}$
                 &  $ s_{\slp} g_1$ \\
\hline
$(\tas_1,\tau)$  &  $-c_{\slp} \f{g_1}{2}$ 
                 &  $ c_{\slp} g_1 $  \\
\hline
$(\tas_2,\mu)$   &  $-s_{\slp} \f{g_1}{2}$
                 &  $-s_{\slp} g_1$ \\
\hline
$(\tas_2,\tau)$  &  $-c_{\slp} \f{g_1}{2}$ 
                 &  $-c_{\slp} g_1$   \\
\hline\hline
\end{tabular}
\end{center}

\bigskip

\bigskip

\vspace*{15mm}
\noindent
{\bf 3.3. Bounds on the LFV parameters from $l_i \to l_j +\gamma$ }
\vspace*{3mm}

Although the slepton mixing is constrained by low-energy data,
it mainly suppress the LFV's associated with the 
first two family sleptons, but still allows the flavor-mixings 
between the second- and third-family sleptons, the smuon ($\mus$) and 
stau ($\tas$), to be as large as $O(1)$~\cite{FCNC}. 
 Here, we are interested in obtaining bounds on the
parameters $x$ and $m_0$, applying the exact slepton 
mass-diagonalization to evaluate the LFV transition 
$\tau \to \mu +\gamma$. In our scheeme, since the selectron 
decouples form the other sleptons, it is not possible 
to induce LFV transitions for the first family.

 Using the interaction lagrangian (15), one can rewrite the general 
expressions for the SUSY contributions to the decays 
$l_i \to l_j +\gamma$ given in Ref. \cite{hisanoetal}.
 The expression for the decay width $\Gamma(\tau \to \mu+\gamma)$,
including the bino-smuon and stau contributions, has the form:
\beq
 \Gamma (\tau \to \mu+\gamma)= \f{\alpha  m^5_\tau}{4\pi}
   [ \sum_\alpha |A_{L\alpha}|^2+|A_{R\alpha}|^2 ]
\eeq
where
\beq
 A_{R\alpha}  = \f{1}{ 32\pi^2 m^2_{{\slp}_\alpha} }
  [\eta^R_{{\slp}_\alpha \tau} \eta^R_{{\slp}_\alpha \mu} f_1(x_\alpha)
          + \eta^R_{{slp}_\alpha \tau} \eta^L_{{\slp}_\alpha \mu}
       \f{m_{\tilde{B}} }{m_\tau} f_2(x_\alpha) ]
\eeq
with $x_\alpha=m^2_{\tilde{B}}/m^2_{{\slp}_{\alpha} } $, and the functions
$f_{1,2}(x_\alpha)$ are given in ref. \cite{hisanoetal}. $A_{L\alpha}$ is
obtained by making the substitutions $L \to R, R\to L$ in Eq. (20).

The decay width depends on the SUSY parameters, and determining the
allowed region in the plane $x-\sm0$ seems the most convenient
choice. Using the current bound $B.R.(\tau \to \mu+\gamma) <  10^{-6}$
gives the exclusion limits shown in fig. 2.
We can see that for values of the scalar mass parameter 
$\sm0 \simeq 200$  GeV, the proportionality solution to the SUSY 
flavor problem is at work, 
while for heavier masses, i.e. $\sm0 \simeq 650$ GeV, it is the
decoupling solution the one that works. For $\sm0 \simeq 600$ GeV, 
and $\tan\beta=5$, $x$ is allowed to be as large as 2.5,
which will induce a large slepton mixing and Higgs-FV transitions,
that could possibly be detectable. 

 On the other hand, SUSY contributions to the the muon anomalous
magnetic moment $\Delta a_\mu$, can also be discussed in this framework,
but  given the present uncertainties regarding the hadronic corrections,
it is enough to consider only the LFV tau decay bound.

\vspace*{5mm}
\noindent
{\bf 3.4. The corrections to $h^0 l_i l_j$ and 
          $h^0 l_i l_i$ Vertex  }
\vspace*{3mm}

Large SUSY correction to the Yukawa couplings, in particular those
related to the top and bottom couplings that arise in the large $\tan\beta$,
have been studied in the literature \cite{hallratsar}, while the
flavor changing Higgs couplings that arise in the u- and d-type
quark sector were studied in \cite{oursqmix} and \cite{babukolda},
respectively. Before describing the full SUSY corrections to the 
Yukawa matrices within our framework,
it is convenient to discuss them using the notation of ref.
\cite{carenaetal}. There, the tree-level lepton Yukawa matrix $h^l_{ij}$
is modified by radiative corrections, but in such a way that the leptons 
couple to both Higgs doublets $H_d$ and $H_u$. Namely,

\beq
{\cal{L}}_Y=  \bar{E}_i [ h^l_{ij} +\delta h^l_{ij} ]H_d L_j +
              \bar{E}_i (\Delta h^l)_{ij} H_u L_j +h.c.
\eeq
where $L_i,E_j$ denote the three-family lepton doublets and singlets, 
respectively.

The soft-brealing terms $A^l_{ij}$ contribute to  $\delta h^l_{ij}$
through the slepton-gaugino loops, thereby realizing the {\it{radiative}}
flavor-Higgs mediation mechanism. One can show that for most regions
of parameter space the corrections to the lepton masses are
proportional to $\cos\beta$ and are thus supressed for large
$\tan\beta$. Moreover, it can be shown that the corresponding
corrections to the diagonal Higgs-lepton couplings are negligible too.
Although one could expect that the FV
Higgs-lepton couplings are also supressed, it turns out that
this is not the case, and the correspoding
LFV Higgs decays, such as $h\to \tau \mu$, are induced at
appreciable rates.
 Furhermore, one does not need to have very large LFV couplings,
since the striking characteristics of the LFV Higgs signals
will facilitate its observation at future colliders.

 On the other hand, the SUSY-conserving (F- and D-terms) do
contribute to $\Delta h^l_{ij}$, and may be enhanced for large
$\tan\beta$. However, one can see that such correction
are dominantly flavor-conserving, because the $H_u-{\slp}_i-{\slp}_i$
couplings are diagonal in flavor space.
Thus, Eq. (18) provides a consistent treatment of the 
radiative flavor-Higgs mediation. Namely, one can use the dominant
correction contained in Eq. (21) to describe the LFC Higgs interactions 
$h\tau^+\tau^-$, as it was done in Ref. \cite{carenaetal}. 
 In fact, their numerical results show that these corrections are 
observable only for a limited region of parameter space.

 An estimate the LFV Higgs couplings $h l_i l_j$, could be
obtained along these lines, using only the approximate
expression for $\delta h_{ij}$. 
However, to take advantage of having a more precise sfermion mass 
diagonalization, we prefer to perform a complete loop calculation, which
allows us to study the effect of the slepton mass-eigenstates in
the loop amplitude.
Thus, we shall write the vertex $h l_i l_j$ as follows:
\beq
g_{hl_il_j}= [(F_L)_{ij} P_L + (F_R)_{ij} P_R ]
\eeq
i.e. in terms of the form factors $F_{L,R}$.
Due to the small lepton masses, one can safely neglect the contribution
from the self-energy corrections (from smuon-bino and stau-bino loops),
and thus $F_{L,R}$ contains only the vertex corrections (from
smuon-stau-bino triangle loops).
\footnote{Additional graphs involving charginos and sneutrinos
are neglected here, mainly to reduce the number of parameters,
which is also equivalent to assume that sneutrinos are much heavier
than charged sleptons.}

In our minimal FC scheme, with large smuon-stau mixing,
the SUSY slepton-bino loops can induce the LFV higgs
decay  $h\to \tau\mu$.
It is known that the branching ratio for this mode,
within the context of the SM with
light neutrinos, is extremely small,
($\lesssim 10^{-7}-10^{-8}$ ) \cite{ourhlfv},
so that this channel becomes an excellent window 
for probing new physics\,\cite{ourhlfv,hanmarfat,marcsher,mysnowtk}.

 Therefore, the one-loop SUSY-EW induced
amplitude for $h^0 \to l_i l_j$ is given by:
\beq
A (h\to l_il_j) \, =\, 
i\,\ov{u}_i(k_2)\(F_LP_L+F_RP_R\)u_j(k_1) \,,\\[4mm]
\label{eq:Hlilj}
\eeq 

The resulting expressions for $F_{L,R}$, including only the
vertex corrections, are:
\beqn
\dis 
F^{V}_{L} & \!\!=\!\! & \frac{g^2_1 m_{\tilde{B}} }{32 \pi^2}                   
         \!\sum_{\alpha\beta}\!
            \lambda^L_{jk} C_0
           (m_h^2,m_{\tau}^2,0; m_{\tilde{l}_{\alpha}},
            m_{\tilde{B}}, m_{\tilde{l}_{\beta}})\,,
\nonumber
\\[-1mm]
\label{eq:FVERch}
\\[-2mm]
\dis
F^V_{R} & \!\!=\!\! & \frac{g^2_1 m_{\tilde{B}} }{32 \pi^2}                  
         \!\sum_{\alpha \beta}\!
            \lambda^R_{\alpha \beta} C_0
           (m_h^2,m_t^2,0; m_{\tilde{l}_{\alpha}},
            m_{\tilde{B}}, m_{\tilde{l}_{\beta}})\,,
\nonumber
\eeqn
where 
$\tilde{l}_{\alpha,\beta}\in (\mus_1,\mus_2,\tas_1.\tas_2)$,
$C_0$ denotes the 3-point $C$-function of Passarino-Veltman.
$\lambda^{L,R}_{\alpha k}$ is the product of the relevant
$h\slp_{\alpha}\slp_{\beta}$ and
$\slp_{\alpha}\tilde{B}\tau(\mu)$ couplings,
i.e.
$\lambda^{L}_{\alpha \beta}= {\cal{H}}_{\alpha \beta}
               \eta^{L}_{\alpha \mu} \eta^{R*}_{\beta \tau}$ 
and
$\lambda^{R}_{\alpha \beta}= {\cal{H}}_{\alpha \beta}
               \eta^{R}_{\alpha \mu} \eta^{L*}_{\beta \tau}$ .

 Finally, the width for the decay procces $h^0 \to l_i l_j$ 
(adding both final states $l^+_i l^-_j$ and $l^-_i l^+_j$ )
is given by:
\beq
\Gamma (h\to l_il_j) \, =\, 
\frac{m_h}{8\pi} (|F_L|2+|F_R|^2) \,,\\[4mm]
\label{eq:Galilj}
\eeq 

On the other hand, if we were interested in using a
lagrangian of type (1) to describe the light Higgs boson,
we would have to work in the decoupling limit, namely when the
remmaining Higgs sector is very heavy ($m_A >> m_Z$), though in fact
this can be achieved even for moderate masses
of order $m_A \gae 600$ GeV.
 Results for the branching ratio of the LFV Higgs mode are shown
in table 2 for several combinations of parameters, which are consistent
with the bounds obained from the LFV tau decay.

\bigskip

\bigskip

\noindent
{Table\,2. 
Br$[ h \to \tau \mu]$
is shown for a sample set of SUSY inputs with 
$(\mu,m_A)=(0.2,0.3)$\,TeV, $A=\f{\sm0}{2}$
and $\tan\beta=5 (10)$.
The numbers in each entry are obtained using the maximum
value $x_{max} (\simeq 1.2-3.0)$ allowed for the
given set of SUSY parameters.}
\vspace*{1.5mm}
\begin{center}
\begin{tabular}{|c|c|c|}
\hline
$m_{\tilde B}$ & $\sm0=450$ GeV & $\sm0=600$ GeV \\ 
\hline
 150 GeV   & $1.1 \times 10^{-7}$ ($3.0 \times 10^{-8}$)
           & $5.0 \times 10^{-5}$ ($1.2 \times 10^{-5}$)  \\
\hline
 300 GeV   & $3.1 \times 10^{-7}$ ($8.0 \times 10^{-8}$)
           & $8.0 \times 10^{-5}$ ($2.1 \times 10^{-5}$)   \\
\hline
 600 GeV   & $5.3 \times 10^{-5}$ ($1.4 \times 10^{-5}$)
           & $4.4 \times 10^{-4}$ ($1.2 \times 10^{-4}$)   \\
\hline
\end{tabular}
\end{center}

\bigskip

\bigskip

In order to study the possibility to detect the LFV higgs decays
at hadron colliders, one can use the gluon-fusion mechanism for 
single Higgs production. Assuming that the production cross-section 
is of similar strength to the SM case, about 1.2 pb for $m_H=125$ GeV 
at Tevatron, it will allow to produce 12,000 Higgs bosons with an 
integrated luminosity of 10 $fb^{-1}$. Thus, for 
$B.R.(H\to \tau \mu/\tau e) \simeq 10^{-1}-10^{-2}$  
Tevatron can produce 1200-120 events.
While at LHC, it will be possible to produce about $10^6$
Higgs bosons through the gluon fusion mechanism \cite{hspira},
with an integrated luminosity of $100$\,fb$^{-1}$.
Then, to determine the detectability of the signal,
 we need to study the main backgrounds to the $h \to \tau\mu$ signal, which
are dominated by Drell-Yan tau pair and WW pair production.
 In Ref.~\cite{hanmarfat} it was proposed to reconstruct the
hadronic and electronic tau decays, assuming the following cuts:
i) For the transverse muon and jet momentum: $p^\mu_T > m_h/5$,
$p^{\pm}_T > 10$ GeV,
ii) Jet rapidity for Tevatron (LHC): $|\eta| < 2 (2.5)$,
iii) The angle between the missing transverse momentum and the
muon direction: $\phi(\mu,\pm)> 160^o$.
The resulting bounds on the LFV higgs couplings, can be expressed 
as a minimum b.r. required to have a $3\sigma$ signal, as shown in
table 3. 

\bigskip

\bigskip

\noindent
{Table\,3. Minimum $B.R.(h \to \mu \tau )$ that can allow detection of the 
LFV Higgs decays. Results are shown for Tevatron Run-2 with 20 (60) 
 $fb^{-1}$, and the LHC with 10 (100) $fb^{-1}$, for $m_h=125$ GeV.}
\vspace*{1.5mm}
\begin{center}
\begin{tabular}{|c|c|c|}
\hline
            &   Run-2  &  LHC  \\
\hline
$B.R._{min}$ & $5.\times 10^{-2}$ ($3.\times 10^{-2}$) 
                             & $5.\times 10^{-3}$ ($8.\times 10^{-4}$) \\
\hline
\end{tabular}
\end{center}

\bigskip

\bigskip

For the  region of parameter space where
$m_{\tilde{B}} \simeq \sm0 \simeq 600 GeV$ and low
$\tan\beta$,  we can obtain $B.R. \simeq 4 \times 10^{-4}$,
which is several orders of magnitude larger than the SM result,
and only about one-half of the value required to get a 
$3\sigma$ signal at LHC \cite{hanmarfat,mysnowtk}. This result  
can be taken as a motivation to look for further improvements in the search 
strategy to discriminate the signal from the SM backgrounds, or to 
consider several running years to enhance the luminosity 
\cite{meandcarmine}.
 These  decay branching ratios are very sensitive to the 
mixing parameter $x$. One reason is that the branching ratio 
(or decay width) contains, besides other mass-diagonalization 
effects, a power factor $x^2$ associated with the Higgs-FV
coupling that appears in the triangle loops.
Another reason is that unlike the usual analyses with mass-insertion
approximation, we have performed exact slepton mass
diagonalization, so that staus and/or smuons can have significant
mass-splittings, as was also shown in Fig. 1.

\vspace*{5mm}
\noindent
{\bf 4. Higgs sector in an $E_6$-inspired  Model
          with a Horizontal $U(1)$ Symmetry} 
\vspace*{3mm}

Multi-Higgs SUSY models are particularly motivated by $E_6$
unification models, where a Higgs pair could be associated with each
family; so to say: each generation requires its own Higgs
sector \cite{sheretal}. If one assumes that the Higgs-Yukawa superpotential
does not permits intergenerational couplings for the Higgs 
superfield, then the phenomenology of the flavor-Higgs sector of the
models is quite simple: there are no FCNC mediated by scalars,
although a reach phenomenology associated with
the multiple Higgs particles will arise.

 On the other hand, one can provide a more
theoretically compelling construction for the
flavor sector of such model, based upon a minimal family symmetry.
 This attractive approach makes use of the simplest
horizontal $U(1)_H$ symmetry to generate a realistic flavor
structure of both fermions and sfermions,
via proper powers of a single suppression factor\,\cite{FN,U1H}.
For convenience, we define the suppression factor
$\ep=\langle S\rangle/\cut$ 
to have a similar size as the Wolfenstein-parameter $\la$ in
the CKM matrix, 
i.e., $\ep\simeq \la \simeq 0.22$\,\cite{U1H}.
Here, $\langle S\rangle$ denotes the vacuum expectation
value of a singlet scalar $S$, responsible for 
spontaneous $U(1)_H$ breaking,
and $\Lambda$ is the scale at which the $U(1)_H$ breaking is 
mediated to light fermions.
In general, the supermultiplets of three-family fermion/sfermions 
may carry different $U(1)_H$ charges.
The Yukawa lagrangian, which is obtained from the superpotential 
of the model, is given by:

\begin{equation}
{\cal{L}}_Y= \bar{U}_i Y^u_{ij} H^u_\alpha Q_j -
             \bar{D}_i Y^d_{ij} H^d_\alpha Q_j -
              \bar{E}_i Y^l_{ij} H^d_\alpha L_j 
\end{equation}
where $H^{u,d}_\alpha$ ($\alpha=1,2,3$) denote the three Higgs pairs
of the model.

 We choose to work in a basis where only $H^{u,d}_3= H_{u,d}$ aquires
a v.e.v.,($<H^0_{u,d}>=v_{u,d}$). Then, assuming that all Higgs
pairs have vanishing charges under the flavor symmetry
$U(1)_F$, we can induce Yukawa couplings that satisfy current
data on quark-masses and CKM angles (which can all be
counted in powers of $\la=0.2$), with the set of $U(1)_H$ quantum 
numbers that appear in Table 4. We are also
considering here $\tanB \sim O(1)$.

\bigskip

\bigskip

\noindent
{Table\,4.\,Quantum number assignments are derived 
with $\tanB \sim O(1)$.  
}
\vspace*{-1mm}
\begin{center}
\begin{tabular}{ccc|ccc|ccc|ccc}
\hline\hline
&&&&&&&&&&&\\[-2.5mm]
$Q_1$      & $Q_2$      & $Q_3$            &   
$~\ov{u}_1$ & $~\ov{u}_2$ & $~\ov{u}_3$    &
$~\ov{d}_1$ & $~\ov{d}_2$ & $~\ov{d}_3$    & 
$~H_a$      & $~H_a$      & $~S$           \\ [1.5mm] 
\hline
&&&&&&&&&&&\\[-2.5mm]
$h_1$        & $h_2$       & $h_3$        &
$\alpha_1$   & $\alpha_2$  & $\alpha_3$   &
$\beta_1$    & $\beta_2$   & $\beta_3$    &
$\xi$        & $\xi'$      & $\gamma$           \\[1.5mm]
\hline
&&&&&&&&&&&\\[-2.5mm]
$4$          & $3$         & $0$          &
$3$          & $0$         & $0$          &
$4$          & $3$         & $3$          &
0        & 0   & $-1$           \\[1.5mm]
\hline\hline
\end{tabular}
\end{center}
\vspace*{3mm}

\bigskip

For instance, the resulting up-quark mass-matrix takes the form of 
\beq
M_u \,\sim\, \dis \f{v_u}{\qs}
\left\lgroup
            \bea{ccc}
            \la^7  &  \la^4  &  \la^4 \\
            \la^6  &  \la^3  &  \la^3 \\
            \la^3  &   1     &   1    
            \eea
\right\rgroup ,
\label{eq:MuAupower}
\eeq
whicih gives the correct spectrum indeed.

On the other hand, since the neutral lepton sector is less constrained,
and only recently the experimental facilities have started to provide
data on the neutrino sector \cite{superkam}, we choose 
to work only with the two-flavor case, namely with the tau and muon
leptons.
Again, working in the basis where only $H^{u,d}_3= H^{u,d}$ 
aquires a v.e.v., the charged lepton mass matrix is: 
$M_l= \f{v_d}{\sqrt{2}} Y^l$.
Then, using the following flavor-symmetry charges 
: $(h_2,h_3)= (2,2)$ and $(\beta_2,\beta_3)= (3,1)$,
for the lepton doublet and singlet, respectively,
we obtain the following charged lepton mass matrix:

\beq
M_l \,\sim\, \dis \f{v_d}{\qs}
\left\lgroup
            \bea{cc}
             \la^5  &  \la^5 \\
              \la^3     &  \la^3  
            \eea
\right\rgroup ,
\label{eq:Mlept}
\eeq
This mass matrix can be diagonalized by a simple 2x2 rotation
matrix parametrized by a mixing angle $\theta_l$, and it can be
verified that this gives the correct order of magnitude for the
charged lepton masses, nameley 
$m_\mu \simeq m_\tau \la^2 \simeq \la^5 v_d$.
Then, the ``Yukawa matrices'' that generate Higgs-lepton
interactions for the remmaining  Higgs doublets $H^d_{1,2}$,
which do not contribute to lepton masses, are given to
leading order by:

\beq
Y^l_{1,2} \,=\, \dis \f{v_d}{\qs}
\left\lgroup
            \bea{cc}
             O(\la^5)           &  O(\la^5) \\
              z_{1,2} \la^3     &   \la^3  
            \eea
\right\rgroup ,
\label{eq:Ylept}
\eeq
We have included the factors $z_{1,2}$ to parametrize the 
$O(1)$ coefficients left undetermined by the FN approach. 
After rotating to the lepton mass-eigenstate basis, 
we obtain the following Higgs-lepton interaction
lagrangian:

\begin{equation}
{\cal{L}}_{int}=  \f{\la^3(1+z_i)}{\sqrt{2}} \bar{\tau} H^d_i \tau
                 + \f{\la^3(z_i-1)}{\sqrt{2}} \bar{\tau} H^d_i \mu +h.c.
\end{equation}
which includes Higgs-FV interactions for tau-mu and Higgs-FC 
interactions for the tau.
Now, these interactions can be transmited to the 
light MSSM-like Higgs
boson of the model, by a mixing mechanism, namely we only
need to assume that the neutral states resulting from
$H^d_1$ and $H^d_2$ mix with MSSM-like CP-even Higgs
state \cite{arandasher}.
 This mixing can be treated as a small perturbation, and one can expect
that it does not affect significantly the remmaining properties of 
the light Higgs boson. Furthermore, we can assume that the lightest
state arising from $H_{1,2}$ dominates this mixing, which can 
be parametrized by another mixing angle $\chi_l$,
in such a way that the LFV Higgs coupling
can be written as:

\begin{equation}
{\cal{L}}_{int}=  
\f{g m_\tau \sin\alpha}{\sqrt{2}m_W \cos\beta}
[-\epsilon_l\f{(1-z_1)}{\sin\alpha} \bar{\tau} \mu
+ (1-\epsilon_l\f{(1+z_1)}{\sin\alpha}) \bar{\tau}  \tau +h.c.] h^0
\end{equation}
where we have substituted $\la^3$ by the appropriate power
of the tau mass.

Finally, to use the same notation as in Eq. (24),
we can write the corresponding expressions for the LFV
form factors ($F_{L,R}$), as follows: 
\beq
F_L= F_R= -\f{g m_\tau \epsilon_l(1-z_1)}{\sqrt{2}m_W \cos\beta}
\eeq

 Then we can analyze the branching ratio that results from this
coupling. For the numerical study, we assume that the 
decay modes of the light Higgs boson include 
$h \to b\bar{b},c\bar{c}, gg,WW^*$. 
Since $z_1$ must be of order unity, we consider
two values $z_1=0.75,0.9$.
As shown in  table 5,  for $z_1=0.75,0.9$ and $\epsilon_l=0.1$,
the decay branching ratio 
${\rm Br}[h\to \tau \mu]$ can be of the order $7 \times 10^{-2}$, 
over a large part of the SUSY parameter space, while for 
$\epsilon_l= 0.05$ the B.r. still can reach values of the order 
$10^{-2}$, especially for large values of $\tan\beta$ 
($\simeq 20$ ),  when the mass of the lightest Higgs boson 
$h^0$ is around $115-120$\,GeV.

 Comparing our results with the minimum B.R. of the LFV Higgs 
$h\to \tau\mu$,  that can be detected at future hadron
colliders (shown in table 3), one can see that there is a
significant region of parameters where such a LFV Higgs signal
can be found. In fact, the rates obtained in this model 
for $z_1=0.75$ are at the reach of Tevatron Run-2. While 
the LHC, can also have a great sensitivity to discover the LFV
decay channel $h\to \tau\mu$ in largest portions of parameter
space, and test the model predictions.
 The future Linear Collider, with a high luminosity, is also expected
to have a good sensitivity to detect this channel \cite{marcsher}.

  On the other hand, this model also predicts corrections to
the Higgs-tau couplings, which can be tested at NLC. Table 5 
shows the resulting deviation of the Higgs width ($h\to \tau^+\tau^-$) 
from the MSSM value, defined as:
\begin{equation}
\Delta \Gamma_{h\tau\tau}= \f{\Gamma_{h_{E_6}\tau\tau} }
                             {\Gamma_{h_{MSSM}\tau\tau} }
\end{equation}
This table shows that $\Delta \Gamma_{h\tau\tau}$ can easily be above 0.08, 
which according to current studies, could be measurable at the
NLC. Furthermore, we notice that the values of 
$\Delta \Gamma_{h\tau\tau}$ obtained for $z_1=0.9$ are slightly
larger than those corresponding to $z_1=0.75$, while the 
LFV Higgs decay $h\to \tau\mu$ is larger for $z_1=0.75$. 
LHC on the other hand, can detect the LFV Higgs signal for both 
values of $z_1$.


\bigskip

\noindent
{Table\,5.  Values of $B.R.(h\to \tau\mu)$ and 
$\Delta \Gamma_{h\tau\tau}$ that arise for
$z_1=0.75,0.9$ and $\epsilon_l=0.1$.
Results in each parenthesis correspond to $\tan\beta= 5,10,20$}
\vspace*{1.5mm}
\begin{center}
\begin{tabular}{|c|c|c|c|}
\hline
 $m_A$    &  $z_1$    & $B.R.(h\to \tau\mu)\times 10^3$  
&  $\Delta \Gamma_{h\tau\tau}$ \\
\hline
 100 GeV  &  0.75      &  (0.19, 0.16, 0.15)
                       &  (0.69, 0.72, 0.74)        \\
          &  0.90      &  (0.03, 0.027, 0.024)
                       &  (0.66, 0.69, 0.71)        \\
\hline
 150 GeV  &  0.75      &  (0.64, 0.17,  0.56)
                       &  (0.44, 0.29,  0.04)        \\
          &  0.90      &  (0.10, 0.27,  0.90)
                       &  (0.40, 0.15,  0.01)        \\
\hline
 200 GeV  &  0.75      &  (1.40, 4.80,  17.0)
                       &  (0.23, 0.03,  0.95)        \\
          &  0.90      &  (0.22, 0.76,  2.70)
                       &  (0.19, 0.07,  1.30)        \\
\hline
 250 GeV  &  0.75      &  (1.90, 7.20,  15.0)
                       &  (0.13, 0.06,   2.0)        \\
          &  0.90      &  (0.31, 1.10,  3.90)
                       &  (0.10, 0.13,  2.60)        \\
\hline
 300 GeV  &  0.75      &  (2.40, 8.80,  29.0)
                       &  (0.09, 0.16,  2.80)        \\
          &  0.90      &  (0.38, 1.40,  4.60)
                       &  (0.05, 0.27,  3.50)        \\
\hline
\end{tabular}
\end{center}

\noindent
{\bf {\large 6. Conclusions  }}\\
\vspace*{1.5mm}

We have shown that a more flavored Higgs boson arises when the flavor
structure encoded in supersymmetric extensions of the
standard model is transmited to the Higgs sector.
The flavor-Higgs transmition mechanism
can have a radiative or mixing origin,
as it is illustrated with several examples,
and can produce interesting Higgs signatures that
can be probed at future high-energy colliders.
In this paper we have focused on the possibility of
testing such flavor-Higgs mediation mechanism
through the LFV Higgs decay $h\to \tau\mu$.

Within the minimal SUSY extension of the SM (MSSM),
the flavor mediation mechanism is of {\it{radiative}} type,
as it is realized through gaugino-sfermion loops, which transmit
the flavor structure of the soft-breaking sector to the
Higgs bosons. In particular, we evaluated the
contributions from the general trilinear A-terms both to the 
Higgs LFV and LFC  vertices.
Our results for the branching ratio of the LFV Higgs mode
$h\to \tau\mu$, give $B.R. \simeq 4 \times 10^{-4}$, which is 
several orders of
magnitude larger than the SM estimate, and about one-half of the
value required to obtain a $3\sigma$ effect at the LHC.
This result is quite motivating to look for further improvements in the
search strategy to discriminate the signal from the SM backgrounds
and to combine several running years to enhance the luminosity.

On the other hand, as an example of flavor mediation through
{\it{mixing}}, we have discussed an $E_6$-inspired multi-Higgs model,
suplemented with an abelian flavor symmetry, where large 
LFV-Higgs effects are also found to arise, though in this 
case at tree-level. We find that even Tevatron can detect the LFV 
Higgs mode $h\to \tau \mu $ for some values of the model parameters.
 While LHC can provide further information on the
flavor structure of the model for other values of such 
parameters. Our results also indicate that deviation from the SM for 
the Higgs-tau-tau coupling could be measurable at the NLC. 

In summary, our results suggest that the Higgs boson that arise
in several well motivated supersymmetric models could have a
more flavored profile, and the future high-energy colliders should
be prepared to allows us to taste it.

\vspace*{4mm}
\noindent
{\bf {\large Acknowledgments}}\\[1.7mm]
I would like thank C.P. Yuan and H.J. He for 
valuable discussions and a very fruitfull collaboration, to Dr. Jaime 
Hernandez and O. Felix for technical assistance and to K. Babu for reading 
the manuscript.
This work was supported by CONACYT and SNI (Mexico), while the submision 
could be made  thanks to the hospitality and support of CERN.




\bigskip

{\bf{LIST OF FIGURES}}

\bigskip

Figure 1. Mass spectrum for the smuon and stau sleptons as a
function of the SUSY scale ($\tilde{m}_0$), for
$x=0.1,0.5$ and $\tanB=5,20, 50$. The line labeled with
d coresponds to the degenerated case.

\bigskip

Figure 2. Allowed values of the parameter $x$ obtained from
$\tau \to \mu+\gamma$, for $\tanB=5$ and bino mass
$m_B=300, 600$ GeV.

\bigskip

Figure 3. Branching ratios for $h\to \tau\mu$ within the $E_6$
inspired model, as function of $\tan\beta$, obtained for 
$z_1=0.75,0.9$ and $\epsilon_l=0.05,0.1$.
The lower line (dots) corresponds to $m_A= 100$ GeV, while
the next ones correspond to: 150 GeV (dot-dot-dash), 200 GeV (dot-dash), 
250 GeV (dots, again), 300 GeV (dashes), while the upper one (solid)
corresponds to 350 GeV.

\end{document}